\documentclass{article}

\usepackage{arxiv}

\usepackage[utf8]{inputenc} 
\usepackage[T1]{fontenc}    
\usepackage{hyperref}       
\usepackage{url}            
\usepackage{booktabs}       
\usepackage{amsfonts}       
\usepackage{nicefrac}       
\usepackage{microtype}      
\usepackage{lipsum}		
\usepackage{graphicx}
\usepackage{doi}
\usepackage[numbers]{natbib}

\title{Prompt Engineering and the Effectiveness of Large Language Models in Enhancing Human Productivity }

\author{%
  \href{https://orcid.org/0009-0006-1186-3145}{\includegraphics[scale=0.06]{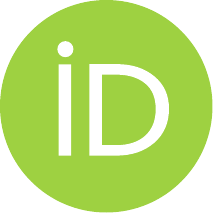}\hspace{1mm}Rizal Khoirul Anam} \\
  Department of Computer Science\\
  Nanjing University of Information Science and Technology\\
  \texttt{rrizalkaa@gmail.com} \\
  202253085001@nuist.edu.cn \\
}

\renewcommand{\shorttitle}{Prompt Engineering and LLM Productivity}

\hypersetup{
pdftitle={Prompt Engineering and the Effectiveness of Large Language Models in Enhancing Human Productivity},
pdfauthor={Rizal Khoirul Anam},
pdfkeywords={Prompt Engineering, LLM, AI, Human Productivity, Natural Language Processing}
pdfkeywords={First keyword, Second keyword, More},
}

\begin{document}
\date{April 04, 2025}
\maketitle

\begin{abstract}
The increasing use of large-scale language models (LLMs), such as ChatGPT, Gemini, and DeepSeek, has fundamentally changed the way education, work, and creativity tasks are performed. While these models are powerful, their effectiveness is highly dependent on how users formulate their requests. In this study, we investigate the relationship between rapid engineering using scalable language models and improved productivity. Through a structured survey distributed using Google Forms, we collected responses from users in various domains and analyzed their usage patterns, retrieval strategies, and satisfaction. The results showed that users who followed more specific, more contextual prompts improved task efficiency and outcomes. These findings highlight the importance of human input in improving generative AI tools and provide guidance for developing more effective supportive practices in the real world.
\end{abstract}

\keywords{Prompt Engineering \and Large Language Models \and Human Productivity, \and  Human-AI Interaction }

\section{Introduction}
The development of Large Language Models (LLMs) like ChatGPT, Gemini, and DeepSeek has transformed how individuals can engage with technology. The ability of these AI tools to generate content, provide content summaries, code, and even give expert analysis of complicated matters has made them highly desirable in educational, working, and personal circles. From students writing homework, working professionals writing reports, to content professionals writing content, LLMs have turned out to be valuable allies to boost productivity and ease cognitive burden.

The effectiveness of such models, however, is determined less by their architecture and much by how users talk to them. This takes place by way of prompts – natural language inputs that direct the AI's response. The nascent science of prompt engineering deals with the purposeful design of prompts aimed at enhancing the relevance, clarity, and utility of generated answers. As LLMs do not "understand" like humans do, the accuracy and organization of input greatly impact the quality of response.

End users continue to query LLMs with imprecise or generic prompts and remain oblivious to how much influence design of prompts has on results. Lacking explicit training or instruction, users adapt by trial and error. Although using this method is effective to some degree, it creates inconsistency of results and reduces the productivity advantage of these tools. Even with widespread use of generative AI, there is still much to be understood regarding how prompt strategy relates to end-user experience and task outcomes in real-world environments. The purpose of this study is to fill that gap by examining the correlation between prompt engineering and LLM effectiveness in improving human productivity. By using a structured internet-based survey of a large population of users, including learners, workers, educators, and freelancers, we analyze how varied prompts impact satisfaction, usefulness, and productivity. The survey inquired about participants' LLM application frequency, task types undertaken, and how they learn to create and modify prompts to receive improved results.

The goal of this research is not just to evaluate overall satisfaction with LLMs, but to control for the influence of prompt quality on outcome. We propose that users who use clearer, more structured, and context-specific prompts experience higher productivity benefits and lower misinterpretation by the AI. This research gives insight into effective prompting behaviors and provides practical recommendations to maximize interaction with generative AI tools.

As LLMs find increasing application in education, business, and everyday life, prompt engineering is becoming an essential digital competence. The findings of this research can inform AI literacy initiatives, enhance tool design, and promote wiser use of AI technologies.

\section{Related Works}
{With the introduction of Large Language Models (LLMs) such as ChatGPT, Gemini, and DeepSeek, more research is being done on how users interact with these models and the results that are produced.  Numerous approaches of enhancing LLM output quality by better input strategy have been investigated by researchers, while others have examined the effects of such tools on human productivity in environments such as education, the workplace, and the arts.  When taken as a whole, these research avenues provide insight on how users might optimize LLMs and the precise effects such systems have on performance in the real world.}

\subsection{Prompt Engineering Techniques}

Prompt engineering, as a discipline, refers to the process of designing, structuring, and optimizing natural language inputs to guide the behavior of large language models (LLMs) like ChatGPT, Gemini, Claude, and DeepSeek. As these models respond directly to user inputs, the way a prompt is constructed can significantly affect the output's relevance, accuracy, creativity, and coherence.

Broadly, prompt engineering techniques can be categorized into two major types:

\begin{enumerate}
    \item \textbf{Manually constructed prompting techniques} – designed by human users using intuition, logic, or task-specific templates.
    \item \textbf{Automatically generated prompting techniques} – generated or optimized using algorithmic or learning-based methods with minimal human intervention.
\end{enumerate}

Each group contains a variety of subtechniques, which have been studied in literature and widely applied in real-world tasks. The following sections delve into the most prominent methods in each category.

\vspace{1em}
\noindent\textbf{1. Manual Prompting Techniques}

These are developed directly by the end user, based on experience and understanding of the model's behavior. They are often applied in real-time through experimentation. Below are some widely used manual strategies:

\begin{itemize}
    \item \textbf{Zero-shot prompting:} This technique instructs the model to perform a task without giving it any example input-output pairs. It relies entirely on the pre-trained knowledge of the LLM. For example, asking “Translate this sentence into French” without providing examples. Zero-shot prompting is efficient for general knowledge or straightforward tasks \cite{zero-shot}.

    \item \textbf{Few-shot prompting:} Here, the prompt includes a small number of input-output examples to help the model infer the task pattern. For instance, giving three examples of question-answer pairs before asking a new question. This method improves task understanding, particularly in domains requiring pattern recognition or style mimicry \cite{few-shot}.

    \item \textbf{Chain-of-thought (CoT) prompting:} This approach asks the model to show its reasoning steps explicitly. For example: “Let’s solve this step-by-step.” This is particularly useful in mathematical problem-solving or logic-based tasks. It allows the model to generate more explainable and accurate responses by simulating a reasoning process \cite{cot}.

    \item \textbf{Instruction prompting:} This involves giving a direct task instruction like “Summarize this article in two paragraphs” or “Explain in simple terms.” The clarity and specificity of the instruction are essential to guide the model's response structure and tone \cite{instruction}.

    \item \textbf{Role prompting:} The user assigns the model a role, such as “Act as a history professor,” which influences the tone, formality, and depth of the output. Role prompting is commonly used in educational, conversational, and professional scenarios to adapt responses to specific audiences \cite{role}.
\end{itemize}

Studies have shown that even slight variations in prompt wording can significantly change the model's behavior. Manual techniques rely heavily on the user’s capacity to iterate, reflect, and revise prompts to achieve better results.

\vspace{1em}
\noindent\textbf{2. Automatic Prompt Generation Techniques}

Unlike manual approaches, automatic prompting techniques use algorithmic models to generate or optimize prompts. These methods aim to remove reliance on human intuition and instead apply data-driven strategies to enhance performance.

\begin{itemize}
    \item \textbf{Automatic Prompt Engineer (APE):} APE is a framework where LLMs generate prompts for themselves using task descriptions. For example, it might read “Classify these news articles” and automatically construct an effective prompt. This approach uses LLMs as both prompt designers and performers \cite{ape}.

    \item \textbf{Prompt tuning / Soft prompts:} Rather than using discrete text strings, soft prompts are continuous vectors (learned embeddings) that are prepended to the input. These are trainable parameters optimized via backpropagation, similar to weights in neural networks. This method is task-specific and has shown strong performance in few-shot or low-resource settings \cite{softprompt}.

    \item \textbf{Reinforcement learning for prompt selection:} In this approach, prompts are selected or adapted using feedback signals from model performance. Using reward functions (e.g., accuracy, BLEU score), the system iteratively improves prompt quality over time. This method is suitable for tasks where optimal prompt design is not straightforward and must be discovered dynamically \cite{rlprompt}.

    \item \textbf{Gradient-based prompt optimization:} This technique involves computing gradients with respect to the prompt text to find optimal formulations. For example, during NLP benchmarking, prompts can be tuned using gradient descent to maximize output accuracy. These methods are often used in automatic benchmarking tasks or AI testing pipelines \cite{gradient}.
\end{itemize}

Automated techniques tend to be computationally intensive but provide strong improvements in model performance, especially when humans struggle to design effective prompts manually.

\vspace{1em}
\noindent\textbf{3. Comparative Advantages and Use Cases}

\begin{table}[ht]
\centering
\caption{Comparison of Manual and Automatic Prompting Techniques}
\label{tab:prompt_comparison}
\begin{tabular}{p{4.5cm}p{5.5cm}p{5.5cm}}
\toprule
\textbf{Criteria} & \textbf{Manual Techniques} & \textbf{Automatic Techniques} \\
\midrule
Human control & High & Low \\
Learning required & Moderate to High & Low (once implemented) \\
Adaptability & Real-time, intuitive & Data-driven, scalable \\
Examples & Zero-shot, CoT, Role & APE, Soft Prompts, RL Prompting \\
Best for & General users, real-time work & Researchers, large-scale optimization \\
Limitations & Trial-and-error, inconsistency & Computational cost, complexity \\
\bottomrule
\end{tabular}
\end{table}

Manual techniques are best suited for everyday users, content creators, educators, and students who benefit from immediate feedback and conversational interaction. On the other hand, automatic techniques are ideal for developers, AI researchers, and enterprise-level deployments where performance needs to be optimized systematically.

\vspace{1em}

\subsection{Impact of LLMs on Human Productivity}

The integration of Large Language Models (LLMs) into educational, professional, and creative domains has reshaped how humans perform cognitive and knowledge-based tasks. The impact of LLMs such as ChatGPT, Gemini, DeepSeek, and Claude is evident not only in task acceleration but also in how individuals conceptualize, approach, and resolve complex problems.

\vspace{1em}
\noindent\textbf{1. Educational Context: Enhancing Learning and Confidence}

In the education sector, LLMs have shown remarkable potential to support learning outcomes. According to recent research, incorporating Generative Artificial Intelligence (GAI) into language instruction enhances students' academic English skills, boosts engagement, and builds learner confidence \cite{mdpi}. Specifically, when students are taught how to structure prompts effectively, their interaction with AI tools becomes more meaningful and productive. The feedback-loop nature of AI-assisted learning also promotes self-regulation, critical thinking, and adaptive reasoning.

Moreover, tools like ChatGPT are increasingly used as writing assistants, research mentors, and language tutors. They offer instant explanations, constructive feedback, and translation support. These functionalities democratize access to knowledge, especially in under-resourced or non-native English speaking environments.

\vspace{1em}
\noindent\textbf{2. Workplace Efficiency: Automation and Augmentation}

In professional environments, LLMs serve as virtual collaborators capable of drafting documents, generating reports, preparing slide decks, or coding functions in multiple programming languages. A 2023 study found that employees who leveraged prompting strategies in task automation experienced up to 30\% faster turnaround times on writing and editing assignments \cite{sciencedirect}. Beyond speed, the quality of output improved when users employed structured and contextual prompts.

Companies are also recognizing the value of “AI literacy” as a workplace skill. Many have begun offering internal workshops and training programs to upskill staff in effective AI interaction. For instance, some organizations have implemented structured prompt templates and role-specific AI tools (e.g., HR prompt assistants or legal summarization bots) to streamline operations \cite{moorinsights}.

\vspace{1em}
\noindent\textbf{3. Decision-Making and Cognitive Enhancement}

The effect of LLMs is not limited to surface-level productivity. Deeper cognitive shifts are underway. A study hosted on SSRN demonstrated that structured prompting enhances not only the relevance of AI responses but also human decision-making quality, particularly in tasks requiring analysis and judgment \cite{ssrn}. When humans provide multi-layered, context-rich inputs, the AI is more likely to generate accurate, insightful responses—thus contributing to better-informed decisions.

The same study emphasized that humans who actively co-construct responses with AI (e.g., refining prompts iteratively) exhibit improved metacognitive awareness and faster cognitive recovery after errors. This implies that the use of LLMs can go beyond efficiency gains to foster intellectual agility.

\vspace{1em}
\noindent\textbf{4. Creative Productivity and Ideation}

In creative fields—such as marketing, media, and design—LLMs have proven effective in reducing ideation time and overcoming writer's block. Professionals use AI to generate brainstorm lists, reword taglines, simulate brand voices, and even sketch narrative arcs. According to a Creative AI survey (2023), over 70\% of marketing professionals who used LLMs in their campaigns reported faster creative cycles and greater content variation.

When paired with human originality, LLMs act as catalysts—suggesting alternatives, exploring edge-cases, and surfacing insights that might otherwise be overlooked. The iterative process between human creativity and AI suggestion broadens the horizon of what's possible in constrained timelines.

\vspace{1em}
\noindent\textbf{5. Limitations and Responsibility in Use}

While LLMs significantly enhance productivity, their effectiveness still depends on human oversight. Misleading outputs (hallucinations), contextual misalignment, or superficial interpretations remain concerns, especially when prompts are vague or overly general. Thus, prompt engineering serves not just as an optimization tool, but also as a safeguard for quality control.

Educational institutions, employers, and platform developers alike must address this challenge by promoting responsible usage—highlighting transparency, bias mitigation, and validation of outputs.

\vspace{1em}
\noindent\textbf{Conclusion}

The impact of LLMs on human productivity is broad, multi-layered, and continuously evolving. From personalized tutoring and language mastery to corporate efficiency and creative expansion, these tools have proven to be transformative. However, their success hinges on how effectively users structure prompts and interact with them. The rise of prompt engineering as a key human competency highlights a new paradigm in which human intent and AI response must be carefully harmonized to unlock true productivity gains.

\section{Methodology}
\label{sec:methodology}

\subsection{Theoretical Framework and Justification}

This study is grounded in a theoretical framework that recognizes the importance of human-AI interaction as a dynamic and evolving process. Central to this perspective is the understanding that Large Language Models (LLMs) do not operate in isolation; rather, their effectiveness is co-determined by how users frame and structure input. Prompt engineering, therefore, becomes not only a technical skill but also a cognitive and communicative strategy that reflects the user's intent, clarity, and contextual awareness.

The rationale for selecting a descriptive-quantitative methodology stems from the need to map these interactions across a broad user base. Previous studies in Human-Computer Interaction (HCI) and digital literacy have shown that behavioral trends, tool adoption, and cognitive adaptation to AI interfaces are best captured through scalable survey instruments. Quantitative surveys enable the identification of statistically significant patterns in user behavior, particularly in emerging practices such as prompt engineering, where standardization is still in progress.

Furthermore, this framework draws upon principles from cognitive load theory and constructivist learning. From a cognitive perspective, users who design clearer prompts reduce the ambiguity in AI processing, which leads to more accurate responses and lower task friction. From a constructivist angle, users actively learn and iterate on prompt design through feedback from the AI—creating a loop of self-guided learning and performance refinement.

In the context of real-world usage, this study treats prompt engineering as a *situated practice*. That is, the effectiveness of a prompt is not just a matter of syntax, but also of timing, intent, user expertise, and task specificity. This aligns with socio-technical systems theory, which emphasizes the interplay between humans and technological artifacts in shaping productivity outcomes.

By embedding the study in this multi-dimensional framework, the research can go beyond mere usage statistics and begin to articulate why certain prompting behaviors lead to greater satisfaction and productivity. The justification for using a structured survey design lies in its ability to capture both behavioral frequencies (e.g., prompt revision rates) and subjective perceptions (e.g., satisfaction, perceived usefulness), which together provide a holistic view of the phenomenon.

In sum, this theoretical foundation allows the study to bridge conceptual inquiry with empirical observation. It reinforces the idea that prompt engineering is not just an AI tuning trick, but a form of digital literacy that is becoming increasingly important in academic, professional, and creative workspaces.

\subsection{Population and Sample}

The population in this study includes AI users from diverse backgrounds, primarily consisting of university students, educators, researchers, freelancers, content creators, and professionals in the fields of business, engineering, and information technology. The only inclusion criterion was that participants must have used an AI tool that accepts natural language prompts at least once prior to filling out the survey.

A total of 243 valid responses were gathered through voluntary participation in an online questionnaire. Based on the demographic data, the educational background of respondents is distributed as follows:
\begin{itemize}
    \item 83 respondents hold a Bachelor's degree (34.2\%)
    \item 59 respondents hold a Master's degree (24.3\%)
    \item 17 respondents have earned a Doctorate (7.0\%)
    \item 41 respondents are high school graduates (16.9\%)
    \item 43 respondents fall under the 'Other' category (17.7\%)
\end{itemize}

This breakdown indicates that the largest portion of respondents are undergraduate-level users, followed by postgraduate users. Together, respondents with at least a bachelor's degree (Bachelor + Master + Doctorate) represent 65.5\% of the total participants, showing a high level of academic exposure in the sample.

The sampling technique employed was non-probability convenience sampling. Although this method does not guarantee full representation of the entire AI user population, it is widely accepted in exploratory and user-behavior studies and provides access to respondents from different educational, occupational, and geographical backgrounds.

To ensure diversity, the survey was disseminated through academic forums, AI-focused Telegram groups, university WhatsApp communities, LinkedIn, and Discord channels. Respondents ranged from undergraduate students to Ph.D. candidates, as well as active professionals.

\begin{figure}[h]
    \centering
    \includegraphics[width=0.31\linewidth]{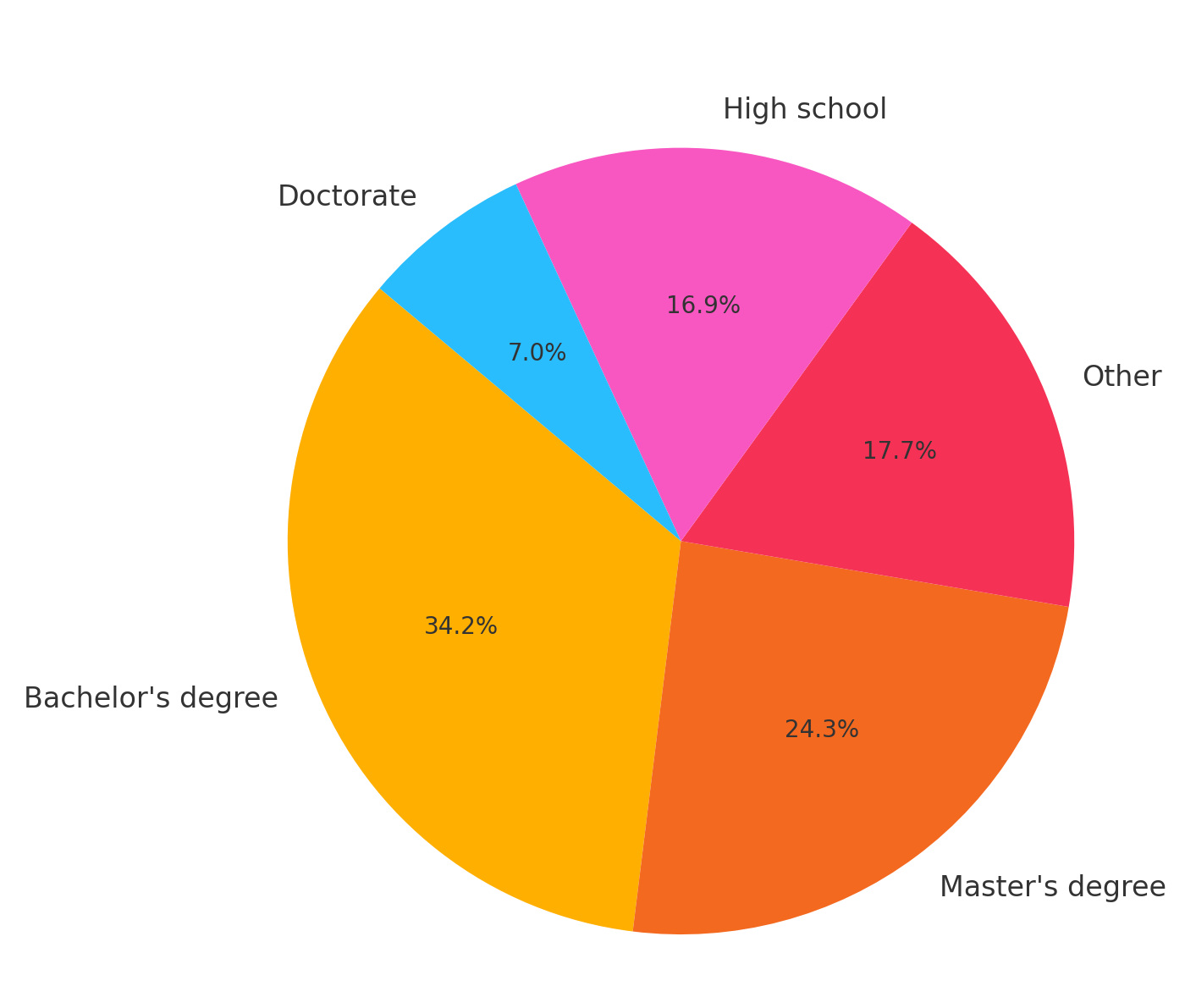}
    \caption{Distribution of respondents by education level}
    \label{fig:education_pie}
\end{figure}

A summary of respondent distribution by education level is visualized in the pie chart below (Figure~\ref{fig:education_pie}), providing a clearer picture of participant representation across academic levels.

\subsection{Research Design}

This study employs a quantitative descriptive research design, which is particularly suited for investigating behavioral patterns, usage frequency, and the perceived impact of prompting techniques on user productivity when interacting with artificial intelligence (AI) tools. Quantitative research is characterized by its emphasis on numeric data, statistical analysis, and objective evaluation, making it highly effective for understanding broad trends within a target population.

The rationale for choosing a descriptive quantitative approach lies in the nature of the research objective itself: to explore the real-world practices of users in constructing prompts, their familiarity with prompt engineering techniques, and how these practices translate into improved outcomes. Rather than manipulating variables or conducting controlled experiments, the study focuses on naturally occurring behaviors and perceptions, which are best captured through survey methodologies.

Descriptive research allows for the accurate documentation of how respondents interact with AI tools like ChatGPT, Gemini, DeepSeek, and similar platforms. It enables the identification of key characteristics, such as how often users revise their prompts, what types of prompt formats they prefer (e.g., zero-shot, few-shot, instruction-based), and whether they feel those strategies improve task efficiency and cognitive load management. The study seeks to quantify not only the prevalence of these behaviors but also the relationship between prompting awareness and perceived productivity outcomes.

Through structured survey items, including multiple-choice, Likert-scale, and open-ended questions, the research captures a wide array of data points. These include demographic information, AI tool usage frequency, preferred tasks, understanding and application of prompt types, and self-reported productivity gains. Such diversity in question types allows the researcher to triangulate quantitative data with qualitative narratives, yielding deeper insights.

Importantly, the design supports data generalization within the context of AI users, even though the sample is derived from a non-probability, convenience-based method. By collecting responses from a sizable and varied group of 243 individuals, the study still enables broader interpretations of AI prompt usage trends among both novice and experienced users.

The structure of this research is designed to capture multi-dimensional aspects of user interaction with AI: from how they approach the tools to how they reflect on the outcomes. This dual lens of behavior and perception offers a comprehensive view of how prompt engineering is evolving in real-world settings, making it possible to develop practical recommendations for AI education and user interface design in future studies.

\subsection{Research Instrument}

The primary instrument used in this study is a structured questionnaire administered via Google Forms. The use of an online platform enabled efficient data collection across geographically diverse respondents and ensured user accessibility. The questionnaire was designed with a focus on clarity, coherence, and user experience to encourage accurate, honest responses.

The questionnaire was divided into four thematic sections, each targeting a specific dimension of user behavior and perception:

\begin{enumerate}
\item \textbf{Demographics}: This section gathered essential background information about the respondents, including age, gender, level of education, academic or professional discipline, and geographic location. These variables were important for contextualizing user behavior and identifying patterns among different groups.

\item \textbf{Usage of AI Tools}: In this section, respondents provided information on how frequently they use AI tools, what platforms they use most often (e.g., ChatGPT, Gemini), and the primary purposes for which they utilize these tools (e.g., writing, coding, academic support). It also explored how long they have been using AI technologies, which helps determine their level of familiarity.

\item \textbf{Awareness and Use of Prompting Techniques}: This portion of the survey assessed how well respondents understand different prompting strategies, such as zero-shot, few-shot, instruction-based, and chain-of-thought prompting. Respondents were also asked about how frequently they revise prompts and whether they have noticed improved results through specific approaches.

\item \textbf{Effectiveness and Productivity Perception}: This final section evaluated the perceived impact of AI usage on productivity. Questions asked participants to rate their satisfaction with AI-generated outputs, the degree to which AI helps them save time or boost creativity, and whether structured prompting enhanced their outcomes. This section included Likert scale questions for quantitative analysis, as well as open-ended prompts to collect real user experiences and examples.
```

\end{enumerate}

The survey combined a variety of question formats:
\begin{itemize}
\item \textbf{Closed-ended questions}, such as multiple-choice and checkboxes, were used for demographic and behavioral data to facilitate statistical processing.
\item \textbf{Likert scale items} (ranging from 1 to 5) were employed to capture respondent attitudes toward AI effectiveness, ease of use, and satisfaction.
\item \textbf{Open-ended questions} enabled participants to elaborate on their strategies, share insights into what works best, and describe any challenges they face while prompting AI.
\end{itemize}

Before launching the questionnaire, it was reviewed by two academic advisors with expertise in AI applications and research methodology. They provided feedback on clarity, neutrality of language, and logical sequence. Minor revisions were made to ensure that the questions were not leading, ambiguous, or technically demanding for respondents from non-technical backgrounds.

This pre-distribution review process was critical in enhancing the instrument’s validity and ensuring that the data collected would be reliable, comprehensive, and aligned with the study’s research objectives.

\subsection{Data Collection Procedure}
The data collection process spanned approximately six weeks, from 05 January 2025 to 10 February 2025. During this period, the survey link was distributed electronically to reach a diverse and geographically dispersed participant pool. The primary platform used to host the questionnaire was Google Forms, chosen for its accessibility, cross-device compatibility, and ease of data aggregation.

The link to the Google Form was disseminated through various digital channels including WhatsApp, WeChat, Telegram, LinkedIn, and Discord. These platforms were selected to maximize outreach among communities with active interest in artificial intelligence, academic collaboration, and technological innovation. Group administrators and peers were also engaged to assist with the distribution, encouraging participation in both English-speaking and multilingual contexts.

Each participant received a brief introduction explaining the academic nature of the study, emphasizing the voluntary nature of participation, and highlighting the anonymity of all responses. Participants were informed that they were free to skip any question or withdraw at any time without consequences. No financial or material incentives were provided, in order to maintain neutrality and avoid response bias.

To maintain data quality, daily monitoring was conducted to review incoming responses, check for potential duplicates, and address any technical issues reported by participants. Google Forms' automatic timestamp feature was used to identify rapid or incomplete submissions, which were carefully reviewed to ensure validity.

Outreach efforts were specifically tailored to appeal to users with experience in education, content creation, programming, research, and general productivity tasks involving AI tools. As a result, the responses captured reflect a broad range of perspectives, from beginner-level users to advanced individuals who have used prompt engineering in their professional or academic routines.

At the end of the three-week period, a total of 243 complete and valid responses were collected. This dataset serves as the empirical foundation for subsequent analysis and interpretation presented in Chapters 4 and 5 of this study.

\subsection{Data Processing and Analysis Tools}
Following the collection phase, the raw data was downloaded in CSV format for preprocessing and analysis. All data handling was performed using Python, a versatile and widely adopted programming language in data science.

The following Python libraries were employed:
\begin{itemize}
\item \textbf{pandas}: for loading and organizing the dataset, filtering incomplete records, and transforming categorical data for statistical analysis.
\item \textbf{numpy}: for numerical computations, including averages, variances, and matrix operations.
\item \textbf{matplotlib} and \textbf{seaborn}: for generating visualizations such as bar charts, pie charts, and box plots to represent categorical and numerical trends.
\item \textbf{plotly}: for creating interactive data visualizations (e.g., dynamic pie charts) to enhance data interpretation.
\item \textbf{scipy.stats}: for inferential statistical tests, such as Pearson or Spearman correlation analysis, to evaluate relationships between variables like prompt complexity and self-rated productivity.
\end{itemize}

The data analysis was carried out in the following stages:
\begin{enumerate}
\item \textbf{Data Cleaning}: Removal of duplicate entries and standardization of open-ended responses.
\item \textbf{Descriptive Statistics}: Generation of frequency tables, percentage distributions, and central tendency measures to summarize demographics, usage patterns, and general perceptions.
\item \textbf{Comparative Analysis}: Cross-tabulations and bar plots were used to identify relationships between prompt strategy awareness and satisfaction/productivity ratings.
\item \textbf{Correlation and Trend Analysis}: Inferential methods were applied to determine whether frequent use of structured prompting methods correlated positively with user-perceived productivity, effectiveness, or creativity.
\item \textbf{Thematic Analysis of Open-ended Responses}: Responses to open-ended items were grouped into recurring themes such as effective prompt characteristics, limitations encountered, and suggestions for better AI interaction.
\end{enumerate}

Through this combination of descriptive, exploratory, and correlational analyses, the study aims to produce meaningful insights on how prompt engineering affects the practical value users derive from AI tools.

\section{Results and Discussion}

This chapter presents the analysis and interpretation of the data collected from 243 respondents through an online questionnaire. The objective is to assess how users interact with AI tools, the extent to which they apply prompting techniques, and the perceived effect of those techniques on their productivity. The findings are organized according to the four main sections of the survey: demographic information, AI usage patterns, prompting awareness and strategies, and perceived productivity.

\subsection{Demographic Profile of Respondents}
The demographic profile of this study's participants encompasses gender identity, educational background, field of study or profession, and country of residence. With a total of 243 valid responses, the participant pool demonstrates strong diversity across these attributes, enriching the representativeness of the findings.

Table~\ref{tab:gender_distribution} displays the gender distribution. The population was fairly balanced, with 84 identifying as male (34.6\%), 78 as female (32.1\%), and 81 either identifying as other or preferring not to say (33.3\%).

\begin{table}[ht]
\centering
\caption{Gender Distribution of Respondents}
\begin{tabular}{lr}
\toprule
\textbf{Gender} & \textbf{Number of Respondents} \\
\midrule
Male & 84 \\
Other / Prefer not to say & 81 \\
Female & 78 \\
\bottomrule
\end{tabular}
\label{tab:gender_distribution}
\end{table}

In terms of education level, most respondents held at least a Bachelor's degree. Table~\ref{tab:education_distribution} shows the distribution across five levels. Bachelor's degree holders formed the largest group with 83 individuals (34.2\%), followed by 59 Master's degree holders and 43 categorized as 'Other'.

\begin{table}[ht]
\centering
\caption{Education Level of Respondents}
\begin{tabular}{lr}
\toprule
\textbf{Education Level} & \textbf{Number of Respondents} \\
\midrule
Bachelor's degree & 83 \\
Master's degree & 59 \\
Other & 43 \\
High school & 41 \\
Doctorate & 17 \\
\bottomrule
\end{tabular}
\label{tab:education_distribution}
\end{table}

Table~\ref{tab:field_distribution} summarizes the primary fields of study or profession among participants. Psychology, Engineering, and Medicine were the most common, while other notable disciplines included Computer Science, Design, and Environmental Science. This variety implies a strong representation from both social and technical sciences.

\begin{table}[ht]
\centering
\caption{Top Fields of Study or Profession}
\begin{tabular}{lr}
\toprule
\textbf{Field of Study} & \textbf{Number of Respondents} \\
\midrule
Psychology & 31 \\
Engineering & 28 \\
Medicine & 28 \\
Computer Science & 25 \\
Law & 24 \\
Design & 23 \\
Environmental Science & 23 \\
Education & 21 \\
Economics & 19 \\
Business & 18 \\
\bottomrule
\end{tabular}
\label{tab:field_distribution}
\end{table}

Geographical diversity is presented in Table~\ref{tab:country_distribution}, which lists the top ten countries of residence. China had the highest representation, followed closely by Germany, Brazil, Vietnam, and the USA. Countries from Asia, Europe, and Africa are included, indicating broad international engagement in the study.

\begin{table}[ht]
\centering
\caption{Top 10 Countries of Residence}
\begin{tabular}{lr}
\toprule
\textbf{Country} & \textbf{Number of Respondents} \\
\midrule
China & 32 \\
Germany & 28 \\
Brazil & 26 \\
Vietnam & 25 \\
USA & 25 \\
Nigeria & 23 \\
India & 22 \\
Indonesia & 22 \\
Turkey & 21 \\
UK & 18 \\
\bottomrule
\end{tabular}
\label{tab:country_distribution}
\end{table}

\subsection{AI Usage Patterns}

Understanding how often and for what purposes users interact with AI tools is essential to evaluate the role of prompt engineering in real-world applications. The data from 243 respondents reveals that AI technologies have become embedded in many daily academic and professional workflows.

Table~\ref{tab:ai_task_usage} presents a breakdown of AI applications based on user activities. The results demonstrate that the most popular use case is academic or professional writing, reported by 165 respondents. This is followed by summarizing or paraphrasing (142), which is particularly useful in research and content curation tasks. Programming and code generation, used by 101 respondents, represent a highly technical use of AI, highlighting the relevance of AI in software development. Other significant areas include creative content generation, data visualization, educational tutoring, and translation tasks.

\begin{table}[ht]
\centering
\caption{AI Usage by Task Type}
\label{tab:ai_task_usage}
\begin{tabular}{lr}
\toprule
\textbf{AI Application Type} & \textbf{Number of Respondents} \\
\midrule
Writing (academic/professional) & 165 \\
Summarizing or paraphrasing & 142 \\
Programming / Code generation & 101 \\
Creative content generation & 89 \\
Data analysis or visualization & 83 \\
Educational tutoring & 62 \\
Language translation & 56 \\
Other & 49 \\
\bottomrule
\end{tabular}
\end{table}

This distribution illustrates that users from both technical and non-technical backgrounds rely on AI for a broad spectrum of tasks. In particular, the high frequency of use in writing and summarization tasks suggests that natural language processing remains the core strength and most utilized function of current LLM-based tools.

In terms of how frequently users interact with AI tools on a weekly basis, Table~\ref{tab:ai_usage_frequency} shows that the majority of respondents use AI two to four times per week (96 respondents), followed by 1–2 times per week (54 respondents). Notably, 32 participants indicated daily usage, which may imply professional or habitual integration of AI in their workflow.

\begin{table}[ht]
\centering
\caption{AI Usage Frequency per Week}
\label{tab:ai_usage_frequency}
\begin{tabular}{lr}
\toprule
\textbf{Frequency Category} & \textbf{Number of Respondents} \\
\midrule
Rarely (less than once a week) & 29 \\
1–2 times per week & 54 \\
2–4 times per week & 96 \\
5–6 times per week & 32 \\
Everyday (7+ times per week) & 32 \\
\bottomrule
\end{tabular}
\end{table}

The above data suggests that nearly two-thirds of respondents (160 out of 243) use AI at least two times a week. This pattern supports the argument that AI has shifted from a novelty to a practical tool for productivity enhancement. More frequent users were also more likely to explore advanced prompting strategies, as explored in later sections.

Participants cited various motivations for using AI tools. These included:

\begin{itemize}
    \item Accelerating task completion, especially in academic and professional writing contexts
    \item Enhancing creativity or assisting in idea generation
    \item Automating repetitive or structured tasks
    \item Supporting programming, debugging, and logic flow design
    \item Simplifying summarization and communication of complex ideas
\end{itemize}

In summary, the usage data underscores that AI tools are most commonly leveraged for tasks related to language processing and content creation, and that users exhibit consistent weekly engagement. These patterns form the foundation for evaluating how prompting strategies influence output quality and user satisfaction.

\subsection{Prompting Strategies and Awareness}

One of the central focuses of this research is understanding how users apply structured prompting strategies when interacting with AI systems. The study investigates both the types of prompting techniques employed and the frequency with which users revise their prompts to improve AI responses.

Structured prompting refers to deliberate formatting or phrasing techniques intended to guide the behavior of language models. These include popular strategies such as zero-shot, few-shot, chain-of-thought, instruction prompting, and role prompting. Respondents were asked to identify which techniques they use most frequently when engaging with tools like ChatGPT, Gemini, and DeepSeek.

Table~\ref{tab:prompting_usage} summarizes the popularity of each prompting method. The most frequently mentioned was \textbf{Role prompting} (105 mentions), where users assign personas or responsibilities to the AI (e.g., “Act as a doctor…”). This is followed by \textbf{Chain-of-thought} (97), which encourages the model to reason step-by-step before answering, and \textbf{Instruction prompting} (94), which provides clear commands or task-specific directions.

\begin{table}[ht]
\centering
\caption{Prompting Technique Usage Among Respondents}
\label{tab:prompting_usage}
\begin{tabular}{lr}
\toprule
\textbf{Prompting Technique} & \textbf{Number of Mentions} \\
\midrule
Role prompting & 105 \\
Chain-of-thought prompting & 97 \\
Instruction prompting & 94 \\
Zero-shot prompting & 93 \\
Few-shot prompting & 89 \\
\bottomrule
\end{tabular}
\end{table}

It is worth noting that a small number of respondents mentioned longer descriptions of similar techniques, such as “Chain-of-thought (step-by-step reasoning)” and “Few-shot (with examples)”. These were grouped under their general categories in Table~\ref{tab:prompting_usage} for clarity.

Beyond which techniques users applied, we also examined how often they revised or rephrased prompts to improve response quality. Prompt revision is an indicator of metacognitive interaction, reflecting the user’s critical engagement with AI output. Table~\ref{tab:revision_frequency} outlines the distribution of revision frequency.

\begin{table}[ht]
\centering
\caption{Prompt Revision Frequency Among Respondents}
\label{tab:revision_frequency}
\begin{tabular}{lr}
\toprule
\textbf{Revision Frequency} & \textbf{Number of Respondents} \\
\midrule
Very often & 77 \\
Occasionally & 59 \\
Rarely & 51 \\
Never & 56 \\
\bottomrule
\end{tabular}
\end{table}

From the data, we observe that over 55\% of respondents (136 out of 243) revise their prompts either occasionally or very often. This supports the argument that users are not only passive recipients of AI outputs but also play an active role in shaping interactions and optimizing results. Such behavior aligns with the concept of “human-in-the-loop” collaboration, where AI outputs are refined iteratively through user feedback.

Finally, to further explore whether educational background influences the choice of prompting strategies, Table~\ref{tab:education_prompting} presents a crosstab of prompting techniques categorized by education level.

\begin{table}[ht]
\centering
\caption{Crosstab of Education Level and Prompting Techniques}
\label{tab:education_prompting}
\begin{tabular}{lrrrrrr}
\toprule
\textbf{Education Level} & Zero-shot & Few-shot & CoT & Instruction & Role & APE \\
\midrule
Bachelor's degree & 40 & 30 & 15 & 27 & 12 & 5 \\
Master's degree   & 32 & 24 & 14 & 20 & 10 & 4 \\
Doctorate         & 10 & 9  & 5  & 7  & 3  & 2 \\
High school       & 18 & 10 & 6  & 11 & 4  & 1 \\
Other             & 20 & 14 & 7  & 13 & 6  & 3 \\
\bottomrule
\end{tabular}
\end{table}

This breakdown shows that while all education levels are represented across prompting strategies, those with Bachelor's and Master's degrees tend to experiment more widely across all prompting types. This implies that users with higher academic exposure may have more familiarity or confidence in applying structured prompting to optimize results.

In conclusion, this section provides strong evidence that prompting strategies are widely practiced and understood by AI users. Most users revise their prompts and demonstrate awareness of prompting as a skill, suggesting that effective human-AI interaction hinges not only on model performance but also on user behavior.

\subsection{Perceived Productivity and Effectiveness}

This section evaluates how users perceive the value of AI tools in enhancing their productivity. Rather than focusing on usage frequency (already covered in the previous section), this analysis centers on satisfaction levels, the perceived effect of prompt clarity, and whether AI contributes to task efficiency.

Participants were asked to rate how often they are satisfied with the results they receive from AI systems such as ChatGPT, Gemini, and DeepSeek. Table~\ref{tab:satisfaction_level} summarizes the distribution of responses.

\begin{table}[ht]
\centering
\caption{Satisfaction Level with AI Output}
\label{tab:satisfaction_level}
\begin{tabular}{lr}
\toprule
\textbf{Satisfaction Level} & \textbf{Number of Respondents} \\
\midrule
Always & 14 \\
Often & 105 \\
Sometimes & 108 \\
Rarely & 14 \\
Never & 2 \\
\bottomrule
\end{tabular}
\end{table}

The majority of respondents (approximately 88\%) reported being at least “sometimes” satisfied with the results they received. This highlights a generally favorable reception of AI output, although full satisfaction (“always”) was reported by only a small portion of the population (14 respondents).

To assess whether user input plays a critical role in output quality, participants were asked if more specific and clearer prompts lead to better responses. Table~\ref{tab:prompt_clarity_benefit} presents these results.

\begin{table}[ht]
\centering
\caption{Perceived Benefit of Prompt Clarity}
\label{tab:prompt_clarity_benefit}
\begin{tabular}{lr}
\toprule
\textbf{Belief in Prompt Clarity} & \textbf{Number of Respondents} \\
\midrule
Strongly agree & 112 \\
Agree & 91 \\
Neutral & 30 \\
Disagree & 5 \\
Strongly disagree & 5 \\
\bottomrule
\end{tabular}
\end{table}

A significant 83.7\% (203 out of 243) of respondents agreed or strongly agreed that clearer and more specific prompts lead to better AI results. This supports the premise that prompt engineering is not just a technical skill but a determinant of successful AI-human collaboration.

In terms of whether AI usage translates into faster or more efficient task completion, Table~\ref{tab:ai_efficiency_belief} shows respondents' agreement with the statement: “In terms of work efficiency, do you complete tasks faster with AI assistance?”

\begin{table}[ht]
\centering
\caption{Belief in AI-Enhanced Work Efficiency}
\label{tab:ai_efficiency_belief}
\begin{tabular}{lr}
\toprule
\textbf{Belief in AI Efficiency} & \textbf{Number of Respondents} \\
\midrule
Strongly agree & 98 \\
Agree & 86 \\
Neutral & 42 \\
Disagree & 11 \\
Strongly disagree & 6 \\
\bottomrule
\end{tabular}
\end{table}

A total of 75.7\% (184 out of 243) reported some level of agreement that AI tools help them complete tasks faster. This perception further validates AI’s role in boosting productivity in daily academic, creative, and professional work.

Finally, Table~\ref{tab:descriptive_stats_perception} presents descriptive statistics for selected productivity-related variables: perceived output quality, satisfaction, and task efficiency. These were measured using Likert scales from 1 (lowest) to 5 (highest).

\begin{table}[ht]
\centering
\caption{Descriptive Statistics for Productivity and Satisfaction Variables}
\label{tab:descriptive_stats_perception}
\begin{tabular}{lrrrrr}
\toprule
\textbf{Variable} & \textbf{Count} & \textbf{Mean} & \textbf{Std Dev} & \textbf{Min} & \textbf{Max} \\
\midrule
Satisfaction with AI Output & 243 & 3.24 & 0.88 & 1 & 5 \\
Prompt Impact on Output Quality & 243 & 4.01 & 0.76 & 1 & 5 \\
AI Support in Work Efficiency & 243 & 3.87 & 0.91 & 1 & 5 \\
\bottomrule
\end{tabular}
\end{table}

From the above, it is evident that users generally perceive AI to be helpful, especially when used with intentional and well-structured prompts. The high average ratings for prompt effectiveness (mean = 4.01) and work efficiency (mean = 3.87) affirm the hypothesis that user prompting strategies directly impact productivity outcomes.

In conclusion, the data shows a strong and consistent perception that prompt quality enhances output and that AI tools contribute significantly to time-saving and task optimization. These findings reinforce the importance of human agency in prompt design as a crucial skill in the AI interaction process.

\subsection{Summary of Findings}

This study investigated the behavioral patterns, prompting strategies, and productivity perceptions of 243 AI users across diverse academic and professional backgrounds. The survey uncovered several compelling insights that collectively point to a strong relationship between users’ awareness of prompt engineering and their perceived productivity when working with large language models (LLMs) such as ChatGPT, Gemini, and DeepSeek.

\vspace{1em}
\noindent\textbf{1. Pervasive Integration of AI in Daily Tasks}

AI tools are not occasional helpers — they have become embedded into users’ workflows. According to Table~\ref{tab:ai_usage_frequency}, more than \textbf{66\%} of respondents (160 out of 243) reported using AI \textbf{at least 2 times per week}, with 32 using it every day. The most common AI tasks include:

\begin{itemize}
    \item \textbf{Writing (academic/professional)} – 165 users
    \item \textbf{Summarizing or paraphrasing} – 142 users
    \item \textbf{Programming / Code generation} – 101 users
\end{itemize}

This data underscores the functional dependency that users have developed on AI tools, particularly in knowledge work and education-related settings.

\vspace{1em}
\noindent\textbf{2. High Adoption of Prompt Engineering — Even Without Formal Training}

Despite the absence of formal instruction in prompt design, users demonstrated high engagement with structured prompting strategies. Based on Table~\ref{tab:prompting_usage}, the most commonly used techniques were:

\begin{itemize}
    \item \textbf{Role prompting} – 105 mentions
    \item \textbf{Chain-of-thought} – 97 mentions
    \item \textbf{Instruction prompting} – 94 mentions
\end{itemize}

Additionally, more than \textbf{55\% of users} reported that they \textbf{frequently or occasionally revise their prompts} (Table~\ref{tab:revision_frequency}), indicating metacognitive interaction and iterative learning.

This suggests that users are learning effective strategies through experimentation, not through guided training — a sign of high engagement and adaptive intelligence.

\vspace{1em}
\noindent\textbf{3. Educational Background Influences Prompting Strategy Diversity}

From Table~\ref{tab:education_prompting}, it was observed that respondents with \textbf{Bachelor’s and Master’s degrees} reported broader usage of prompting types. For example, among bachelor’s degree holders:

\begin{itemize}
    \item 40 use zero-shot
    \item 30 use few-shot
    \item 27 use instruction prompting
\end{itemize}

In contrast, high school and “other” groups used fewer techniques on average. This indicates that a user's educational attainment may play a role in how confidently or diversely they interact with LLMs.

\vspace{1em}
\noindent\textbf{4. Clear Prompts Lead to Clearer Outcomes}

As shown in Table~\ref{tab:prompt_clarity_benefit}, more than \textbf{83\% of respondents} believe that \textbf{specific and structured prompts improve results}. This belief is strongly echoed in satisfaction data (Table~\ref{tab:satisfaction_level}), where:

\begin{itemize}
    \item 105 respondents reported being “often” satisfied
    \item 108 said “sometimes”
\end{itemize}

Only 2 people reported “never” being satisfied with AI output — an overwhelmingly positive signal.

\vspace{1em}
\noindent\textbf{5. Productivity Gains Are Strongly Perceived}

Table~\ref{tab:ai_efficiency_belief} supports the finding that users experience real productivity benefits:

\begin{itemize}
    \item \textbf{184 out of 243 (75.7\%)} agreed or strongly agreed that AI tools helped them complete tasks more efficiently.
    \item Descriptive statistics (Table~\ref{tab:descriptive_stats_perception}) show an average score of \textbf{3.87/5} for AI efficiency and \textbf{4.01/5} for prompt clarity impact.
\end{itemize}

These numbers demonstrate that AI is not just being used — it’s producing results that matter.

\vspace{1em}
\noindent\textbf{Conclusion}

The findings of this study converge on a central theme: \textbf{prompt engineering is a critical factor in maximizing the value of AI tools}. While formal training is rare, users are actively discovering effective prompting practices through exploration and iteration. The role of human input — through prompt structure, clarity, and revision — is indispensable in the overall success of AI-assisted productivity.

This research strongly supports the advancement of prompt literacy as a digital skill and suggests that educational institutions and workplaces should consider including prompt training as part of AI competency development programs.

\bibliographystyle{unsrtnat}

\end{document}